\documentclass{elsart}

\usepackage{natbib}
\usepackage[dvips]{graphicx}
\usepackage{amssymb}

\begin{document}

\begin{frontmatter}

\title{Decision table for classifying point sources based on FIRST and 2MASS databases}

\author{Yanxia Zhang, Yongheng Zhao \and Dan Gao}

\address[label1]{National Astronomical
Observatories, CAS, 20A Datun Road, Chaoyang District, Bejing 100012
China}

\begin{abstract}
With the availability of multiwavelength, multiscale and multiepoch
astronomical catalogues, the number of features to describe
astronomical objects has increases. The better features we select to
classify objects, the higher the classification accuracy is. In this
paper, we have used data sets of stars and quasars from near
infrared band and radio band. Then best-first search method was
applied to select features. For the data with selected features, the
algorithm of decision table was implemented. The classification
accuracy is more than 95.9\%. As a result, the feature selection
method improves the effectiveness and efficiency of the
classification method. Moreover the result shows that decision table
is robust and effective for discrimination of celestial objects and
used for preselecting quasar candidates for large survey projects.
\end{abstract}

\begin{keyword}
 techniques: miscellaneous; methods: statistical; methods: data analysis; astronomical data bases: miscellaneous;
 catalogs; feature selection


\end{keyword}

\end{frontmatter}

\section{Introduction}
With the development of various multiwavelength projects, such as
SDSS, GALEX, 2MASS, GSC-2, POSS2, RASS, FIRST and DENIS, astronomy
is about to undergo a major paradigm shift. Data volumes are
doubling every 20 months. Data sets are becoming larger, and more
homogeneous. It is a challenge to deal with multi-terabyte databases
efficiently and effectively for astronomers. Under this situation,
astronomers will have to be just as familiar with mining data as
with observing on telescopes. Classification, as one of data mining
tasks, is a key issue in astronomy. Celestial objects are divided
into different kinds of objects (e.g. stars, galaxies and quasars)
by spectra, photometry or image, moreover each kind may be further
subdivided.

In the last years there were many data mining algorithms
successfully applied in astronomy. For instance, neural network
methods were used for spectra classification; support vector
machines (SVM) were employed in classification of multiwavelength
data (Zhang \& Zhao 2004); decision trees were used to automatically
classify objects (Jarrett et al. 2000; Ball et al. 2006). In this
paper we discuss an example in which we classify objects as quasars
or stars using the cross-match results between a radio survey (the
Faint Images of the Radio Sky at Twenty centimeters, FIRST) and a
near infrared survey (the Two Micron All Sky Survey, 2MASS) by
decision tables. Based on FIRST and 2MASS databases, the source
candidates selected by decision tables are radio loud objects which
is bright enough in near infrared band to be detected in 2MASS.
According to special issues, astronomers may choose data from
different bands. For example, in order to obtain X-ray strong
objects, the data from RASS may be employed; for the study of
properties of various objects in five optical bandpasses, SDSS is a
good choice. Decision tables, like decision trees or neural
networks, are classification models used for prediction. They are
induced by machine learning algorithms. The classifier trained by
the method helps guide the choice of which objects to follow up with
spectroscopic measurements. Therefore the efficiency of telescopes
will be improved and human efforts will be reduced.

This paper is organized as follows: Section 2 describes the sample
and chosen attributes. Section 3 introduces the principle of
decision tables. Section 4 lists the experiment result and
discussion. Section 5 summarized this work.

\section{Data Sample and Chosen Attributes}
We describe here near infrared, radio and optical catalogs as
follows:

The Two Micron All Sky Survey (2MASS) project (Cutri et al.
2003) is designed to close the gap between our current technical
capability and our knowledge of the near-infrared sky. 2MASS uses
two new, highly-automated 1.3-m telescopes, one at Mt. Hopkins, AZ,
and one at CTIO, Chile. Each telescope is equipped with a
three-channel camera, each channel consisting of a 256x256 array of
HgCdTe detectors, capable of observing the sky simultaneously at j
(1.25~$\mu$m), h (1.65~$\mu$m), and k$_{\rm s}$ (2.17~$\mu$m), to a
3$\sigma$ limiting sensitivity of 17.1, 16.4 and 15.3~mag in the
three bands. The number of 2MASS point sources adds up to
470,992,970.

The Faint Images of the Radio Sky at Twenty centimeters (FIRST)
began in 1993. It uses the VLA (Very Large Array, a facility of the
National Radio Astronomical Observatory (NRAO)) at a frequency of
1.4GHz, and it is slated to 10,000 square degree of the North and
South Galactic Caps, to a sensitivity of about 1mJy with an angular
resolution of about 5 arcsec. The images produced by an automated
mapping pipeline have pixels of 1.8 arcsec, a typical rms of
0.15~mJy, and a resolution of 5 arcsec; the images are available on
the Internet (see the FIRST home page at http://sundog.stsci.edu/
for details). The source catalogue is derived from the images. A new
catalog (Becker et al. 2003) of the FIRST Survey has been released
that includes all data taken from 1993 through September 2002, and
contains about 811,000 sources covering 8,422 square degrees in the
North Galactic cap and 611 square degrees in the South Galactic cap.
The new catalog and images are accessible via the FIRST Search
Engine and the FIRST Cutout Server.

The 12th edition catalogue of quasars and active nuclei (Cat.
VII/248, V\'eron-Cetty \& V\'eron 2006) is an update of the previous
versions, which now contains 85221 quasars, 1122 BL Lac objects and
21737 active galaxies (including 9628 Seyfert 1s), almost doubling
the number listed in the 11th edition. Just like the previous
editions, no information about absorption lines of X-ray properties
are given, but absolute magnitudes are given, assuming
H$_0=50$~km~s$^{-1}$~Mpc$^{-1}$ and q$_0=0$. In this edition the
20~cm radio flux is listed when available, in place of the 11~cm
flux.

The Tycho-2 Catalogue (Cat. I/259, Hog et al. 2000) is an
astrometric reference catalogue containing positions and proper
motions as well as two-color photometric data for the 2.5 million
brightest stars in the sky. The Tycho-2 positions and magnitudes are
based on precisely the same observations as the original Tycho
Catalogue (hereafter Tycho-1; see Cat. I/239) collected by the star
mapper of the ESA Hipparcos satellite, but Tycho-2 is much bigger
and slightly more precise, owing to a more advanced reduction
technique. Components of double stars with separations down to
0.8~arcsec are included. Proper motions precise to about
2.5~mas~yr$^{-1}$ are given.

We obtained 153135 entries with one to one matching between the
FIRST and 2MASS catalogues within 5 arcsec radius. The entries were
then cross-identified with the V\'eron-Cetty \& V\'eron 2006 catalog
and the Tycho-2 catalog within 5 arcsec radius, respectively.
Similarly, we obtained 2389 quasars and 1353 stars from the 2MASS
and FIRST catalogues. The chosen attributes from different bands are
$logF{\rm peak}$ ($F{\rm peak}$: peak flux density at 1.4\,GHz),
$logF{\rm int}$ ($F{\rm int}$: integrated flux density at 1.4\,GHz),
$f{\rm maj}$ (fitted major axis before deconvolution), $f{\rm min}$
(fitted minor axis before deconvolution), $f{\rm pa}$ (fitted
position angle before deconvolution), $j-h$ (near infrared index),
$h-k$ (near infrared index), $k+2.5log{\rm Fint}$, $k+2.5logF{\rm
peak}$, $j+2.5logF{\rm peak}$, $j+2.5logF{\rm int}$, $b-v$ (optical
index). $b-v$ is from the two catalogues: the quasar catalogue of
V\'eron 2006 and the Tycho-2 Catalogue. Since the quasar catalogue
of V\'eron 2006 is an inhomogeneous compilation, and the photometry
from the Tycho-2 Catalogue was made in a specific system whose
conversion to johnson $b-v$ relies on physical assumptions, the
classification regarding $b-v$ is only used as a rough reference.

Zhang \& Zhao (2007) showed that $logF{\rm peak}$, $logF{\rm int}$,
$k+2.5logF{\rm int}$, $k+2.5logF{\rm peak}$, $j+2.5logF{\rm peak}$,
$j+2.5logF{\rm int}$ are useful to classify quasars from stars, and
$f{\rm maj}$, $f{\rm min}$ and $f{\rm pa}$ are unimportant. To
further see the statistical distribution of this sample and compare
the distribution of this sample with all the sources from the
cross-identification of 2MASS and FIRST catalogues, the scatter
plots of some parameters are shown in Fig. 1. The scatter plots also
indicate that the conclusion is reasonable, moreover, $b-v$ is
helpful to nearly completely discriminate quasars from stars.

  \begin{figure}
   \begin{center}
\includegraphics[bb=3 2 391 427,width=10cm,clip]{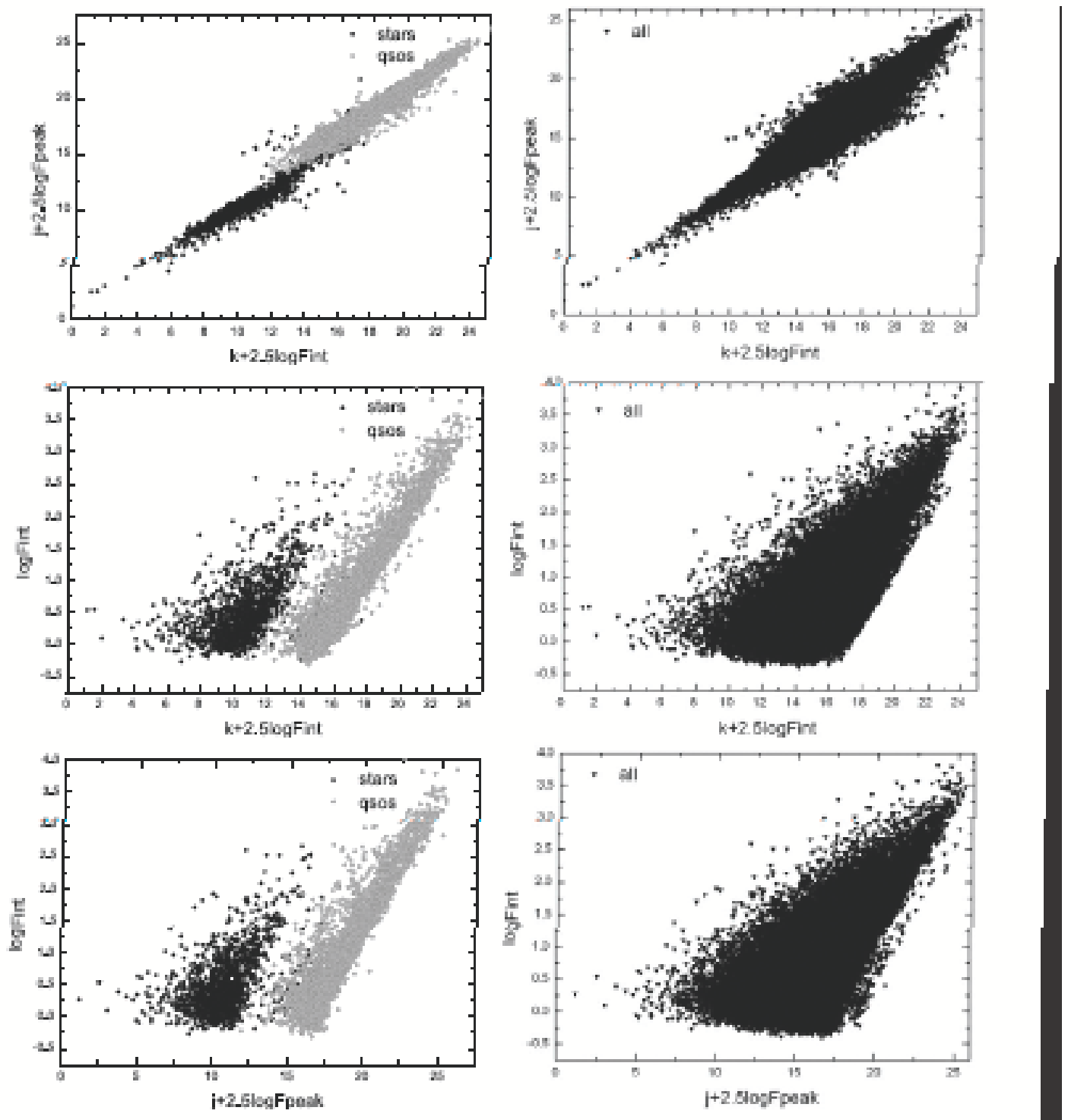}
\includegraphics[bb=3 2 391 427,width=10cm,clip]{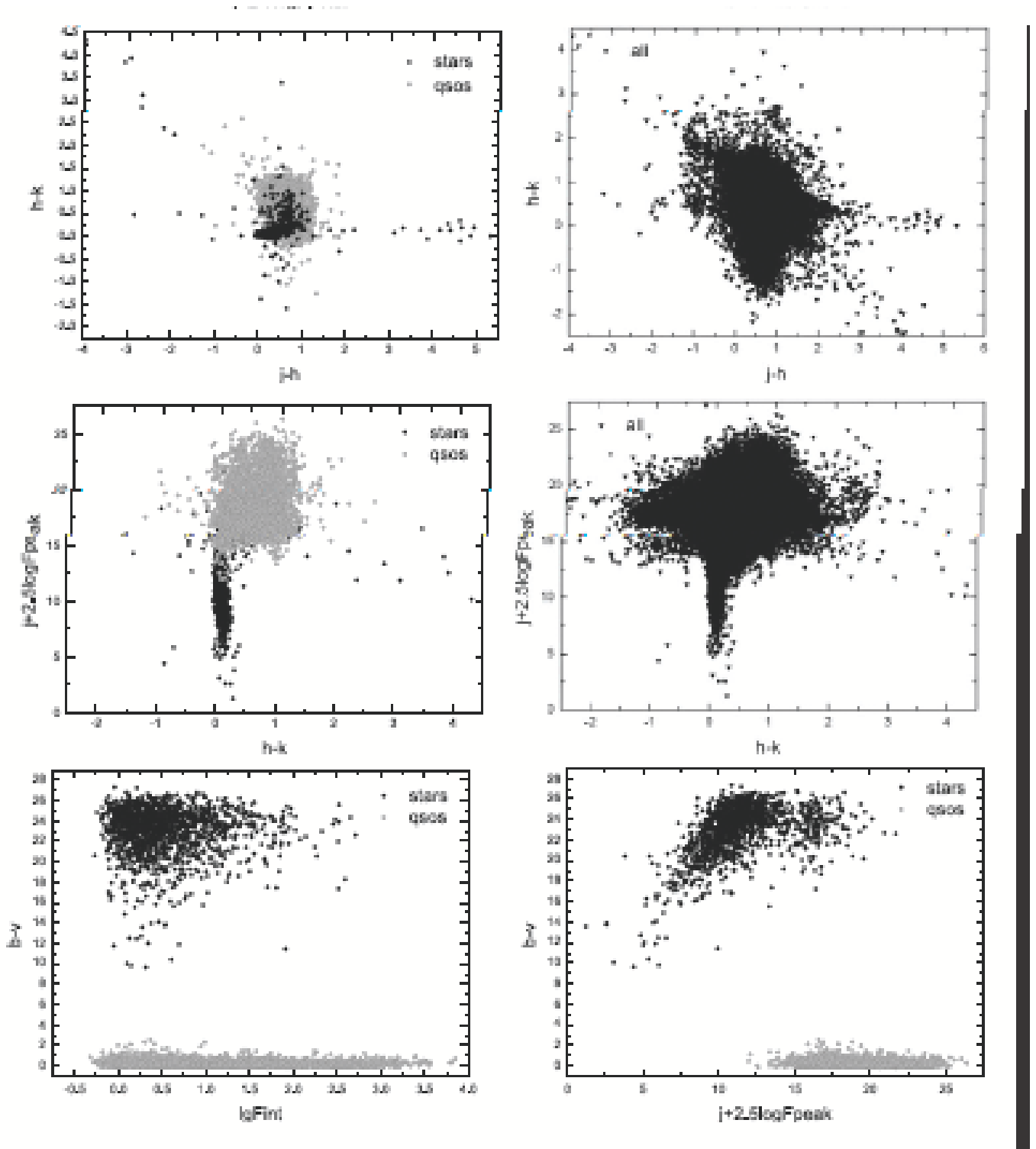}
   \end{center}
   \caption
   { \label{fig:example}
 The scatter plots (filled circles represent stars; open ones
represent quasars; triangles represent all sources from the
cross-identification of 2MASS and FIRST catalogues.) }
   \end{figure}

\section{Decision Tables}
A decision table consists of a hierarchical table in which each
entry in a higher level table gets broken down by the values of a
pair of additional attributes to form another table. The structure
is similar to dimensional stacking. For the detailed principle of
decision table, readers can refer to Kohavi (1995).

Given a training sample containing labelled instances, an
induction algorithm builds a hypothesis in some representation.
The representation we investigate here is a decision table with a
default rule mapping to the majority class, which we abbreviate as
{\bf DTM}. A DTM consists of two components:

1. A {\bf schema}, which is a set of features.

2. A {\bf body}, which is a multiset of labelled instances. Each
instance is made up of a value for each of the features in the
schema and a value for the label.

Given an unlabelled instance $I$, the label assigned to the
instance by a DTM classifier is computed as follows. Let $\ell$ be
the set of labelled instances in the DTM exactly matching the
giving instance $I$, where only the features in the schema are
required to match and all other features are ignored. If
$\ell=\emptyset$, return the majority class in the DTM; otherwise,
return the majority class in $\ell$. Unknown values are treated as
distinct values in the matching process.

Let {\bf err(h,f)} denote the error of a hypothesis $h$ for a
given target function $f$. Since $f$ is never known for real-world
problems, we estimate the error using an independent test set
$\tau$ as

$$
\widehat{err}(h,\tau)=\frac{1}{| \tau |}\sum\limits_{(x_i, y_i)\in
\tau} L(h(x_i),y_i)
$$
where $L$ is a loss function. In the rest of the paper we assume a
zero-one loss function, i.e., zero if $h(x)=y$ and one otherwise.
The approximate accuracy is defined as $1-\widehat{err}(h,\tau)$.

An optimal feature subset, $A^\ast$, for a given hypothesis space
$H$ and a target function $f$ is a subset of the features $A^\ast$
such that there exists a hypothesis $h$ in $H$ using only features
in $A^\ast$ and having the lowest possible error with respect to
the target function $f$. (Note that the subset need not be
unique.) As the following example shows, relevant features are not
necessarily included in the optimal subset.

An induction algorithm using DTMs as the underlying hypothesis
space must decide which instances to store in the table and which
features to include in the schema. The algorithm is assumed to
include the projections of all instances defined by the schema in
the DTM, but we do not restrict the subset of features to use in
the schema in any way. Let $A^{\ast}=\{X_1,...,X_n\}$ be a set of
features and let $S$ be a sample of $m$ instances over the
features in $A$. Given a subset of features $A^{\prime} \subseteq
A$, DTM$(A^{\prime},S)$ is the DTM with schema $A^\prime$ and a
body consisting of all instances in $S$ projected on $A^\prime$.
The goal of the induction algorithm is to choose a schema $A^\ast$
such that

\begin{equation}
A^{\ast}={\arg\min \limits_{A^{\prime}\subseteq A}}
err(DTM(A^{\prime},S),f).
\end{equation}

The schema $A^{\ast}$ consists of an optimal feature subset for a
DTM under the assumption that all instances from the training set
are stored in the body of the decision table.

\section{Best-first Search for Feature Selection}

Both filter and wrapper approaches can be applied for feature
subset selection. Filter approaches use only the training data in
the process of evaluation but wrapper approaches incorporated the
induction algorithm as part of the evaluation in the search of the
best possible feature subset. In this paper we apply best-first
search as wrapping around decision table method to obtain optimal
feature subsets. In order to search the space of feature subsets
effectively, we transform the problem into a state space search
and use best-first search to heuristically search the space. A
forward selection procedure using best-first search is adopted
(Ginsberg 1993). Forward selection implies an operation of
addition for each expansion. The search states are nodes
representing subsets of features. The idea of best-first search is
to jump to the most promising node generated so far that has not
been expanded. The search is stopped when an improved node has not
been found in the previous $k$ expansions. An improved node is
defined as a node that has an accuracy of not less than $x$
percent higher than the best node found so far.

To estimate future prediction accuracy, cross-validation, a standard
accuracy estimation technique (Weiss \& Kulikowski 1991; Breiman et
al. 1984; Stone 1974), is adopted. Given an induction algorithm and
a dataset, $k$-fold cross-validation divides the data into $k$
approximately equally sized subsets, or folds. The induction
algorithm is executed $k$ times; each time it is trained on $k-1$
folds and the generated hypothesis is tested on the rest fold, which
serves as a test set. The estimated accuracy is computed as the
average over the $k$ test sets. If $k$ equals the sample size, this
is called ``leave-one-out" cross-validation. ``Leave-v-out" is a
more elaborate and expensive version of cross-validation that
involves leaving out all possible subsets of $v$ cases.

\section{Experiment and Discussion}
Our experiments were done with the WEKA machine learning package
(Witten \& Frank 2005), which is a collection of machine learning
algorithms for data mining tasks. Now we applied decision table
method on all the datasets including 2389 quasars and 1353 stars.
Best-first search for feature selection was executed and terminated
by leave-one-out cross-validation after 5 non improving subsets. We
considered two situations: the sample with $b-v$ and the sample
without $b-v$. For the sample with $b-v$, the optimal feature subset
was $b-v$ from the 12 features ($logF{\rm peak}$, $logF{\rm int}$,
$j-h$, $h-k$, $k+2.5logF{\rm int}$, $k+2.5logF{\rm peak}$,
$j+2.5logF{\rm peak}$, $j+2.5logF{\rm int}$, $f{\rm maj}$, $f{\rm
min}$, $f{\rm pa}$, $b-v$). Then $b-v$ was used to create a
classifier by means of decision table algorithm. The estimated
accuracy for each node was computed using 10-fold cross-validation.
We got 2 classification rules. The time taken to build the
classifier was 0.69 seconds (the configuration of the personal
computer used to carry out this analysis is Microsoft Windows XP,
Pentium (R) 4, 3.2~GHz CPU, 1.00~GB memory). The whole accuracy
added up to 100\%.

Then given the sample without $b-v$, we obtained the optimal feature
subsets $lgF{\rm int}$, $j+2.5lgF{\rm peak}$ and $k+2.5lgF{\rm int}$
from the 11 features. The number of classification rules is 51. The
time taken to the built model spent 0.66 seconds. Correctly
classified instances were 3558, which occupied 95.1\% of the whole
sample; incorrectly classified instances were 184, occupying 4.9\%.
The accuracy of stars and quasars was 88.0\% and 99.0\%,
respectively.

For the two samples, the classification results were shown in
Table~1, the whole accuracy added up to 100.0\% and 95.1\%,
separately.

\begin{table*}[h!]
\begin{center}
\caption{The classification result with different samples}
\bigskip
\begin{tabular}{r|ll|ll}
\hline \hline
Sample&with $b-v$&&without $b-v$& \\
\hline
classified$\downarrow$known$\to$& stars &quasars& stars &quasars\\
\hline
        stars  & 1353 & 0  &1193&24\\
       quasars & 0    &2389&160 &2365\\
\hline
       Accuracy & 100.0\% & 100.0\%&88.0\%&99.0\% \\
\hline
\end{tabular}
\bigskip
\end{center}
\end{table*}

In our case, when regarding the dataset from three bands, $b-v$ is
taken as the optimal feature by best-first search, which is
consistent with the information from the scatter plot. While
considering the dataset from radio and near-infrared bands, $lgF{\rm
int}$, $j+2.5lgF{\rm peak}$ and $k+2.5lgF{\rm int}$ are selected as
the optimal feature subsets. This shows that the best-first search
is a more effective feature selection technique than histogram of
Zhang \& Zhao (2007). The accuracy (95.1\%) is satisfying, which is
comparable to the accuracy (94.36\%, 95.80\% and 95.19\%) by BBN,
MLP and ADTree, respectively. This result indicates that decision
table method is an efficient and effective algorithm to classify
quasars from stars with the multiwavelength data. The accuracy
(99.0\%) of quasars is higher than that (88.0\%) of stars, which
possibly results from the fact that the number of quasars is larger
than that of stars. Usually, the imbalanced sample is a factor that
influences the performance of a classifier. Classifiers are easy to
remember the rule of the majority of sample. Miss-classified
instances also result from the attribute errors of stars and
quasars. The existence of errors lead to overlap of stars and
quasars in the classification space. Based on FIRST and 2MASS
databases to preselect quasar candidates, the classification rules
are extracted from radio and infrared bands, thus the selected
candidates own characteristics of the two bands. As a result, the
quasar candidates by the classifier are generally radio-loud and red
quasars. If we want to obtain a complete sample, we need to consider
other selection criteria, for example, consider more bands, or
change bands.

\section{Conclusion}

The construction of a complete sample of quasars is helpful for
studying the large-scale structure of the universe and the formation
and evolution of galaxies. To reach this aim, we need efficient
separation of quasars from other astronomical sources, of which it
is an important issue to separate quasars from stars because they
are both point sources from their images. In this work, we focus on
this issue. With the development of detectors and the construction
of observational stations, the observational attributes of celestial
objects increase to hundreds or even thousands. Thus how to reduce
the dimension of data seems more important for data analysts,
astronomers or for the requirement of some algorithms. We applied a
best-first search method to select out the optimal feature subsets.
Then the decision table algorithm was executed on the sample with
$b-v$ and the sample without $b-v$. The classified accuracy is more
than 95.0\%, and the speed to build the model is very high.
Therefore, from the point of view from accuracy and speed, this
classification algorithm is satisfactory, especially faced with huge
volumes and complexity of data. Since the selection criteria
seriously depend on algorithms and data used, the classifiers
obtained in some situation are inclined to select some kind of
quasars. In our case, the classification rules obtained by decision
tables is reasonable and applicable. As shown in Fig. 1, the rules
may be used to preselect radio loud and red quasar candidates from
the whole FIRST and 2MASS intersection. On account of this, we need
to consider more criteria and other data to preselect quasar
candidates. Decision table can be used to construct classifiers with
other datasets and then preselect quasar candidates. Certainly,
according to the interests and issues of astronomers, other kinds of
data may be collected and used as training set, for example,
spectra, images, photometry and so on. For the study of stars,
various star samples are needed. For the study of galaxy, different
galaxy samples are required. The training sample is as complete as
possible and thus the obtained classifier can have a better
prediction ability. Owing to the complex characteristics of
astronomical data, for different problems, we need to develop
appropriate and effective algorithms to solve them.

{\bf Acknowledgments} We are very grateful to anonymous referees for
their helpful comments and suggestions. This paper is funded by
National Natural Science Foundation of China under grant No.10473013
and No.90412016.

\end{document}